\documentclass[nofootinbib]{revtex4-1}
\usepackage{graphicx}
\usepackage[colorinlistoftodos]{todonotes}
\usepackage{dcolumn}
\usepackage{bm}

\begin{document}

\title{Spawning Nodes Generate Deterministic Scale-Free Networks}

\author{Peter R. Conwell}
 \altaffiliation[]{Emeritus Professor, Physics, Westminster College, Salt Lake City, UT, peterrconwell@gmail.com}
\author{Kaushik Chakram}
\author{Valeria J. Villegas-Medina}

\date{5/25/2013}

\begin{abstract}
In this paper we present a deterministic vertex spawning model that yields a scale-free network. The model specifies that a parent vertex produces a child vertex in a time interval approximately proportional to the current time and inversely proportional to the number of edges currently connected to the parent. Spawned offspring maintain an undirected edge with its parent. No information about the network as a whole is required to obtain scale-invariant behavior. Although the algorithm is deterministic, the number of nodes spawning in a small time interval quickly becomes randomized.  We show theoretically and with simulations that such a spawned network will have a degree distribution obeying a power law with exponent 2.5. Simulations show that the distribution matches a Zipf distribution.
\end{abstract}
\maketitle
\section{\label{intro}Introduction}

 Viewed through a certain lens almost any aggregate object, physical or abstract, can be seen as a network of simple nodes (vertices) and the connections between them (edges). Such a network perspective can be a propitious tool to describe generalized topology, growth, and aggregation of disparate entities ranging from the organization of the internet to social interactions. Indeed, network research has become a discipline in and of itself. \cite{Barabasi2016NetworkScience},\cite{dorogovtsev2013evolution},\cite{mej2010networks}

Albert-L{\'a}szl{\'o} Barab\'asi and R{\'e}ka Albert seemingly initiated a phase transition in network research with their 1999 $Science$ article \cite{barabasi1999emergence}. They re-discovered that networks evolving with new nodes attaching preferentially to existing nodes obey a power law distribution
\begin{equation}
\label{PowerLaw}
 p_k=Ck^{-\alpha},  
\end{equation}
where $p_k$ is the probability of a node having $k$ connections, $C$ is a normalization constant, and $\alpha = 3$. Networks that have a power law distribution are said to be scale-free or scale-invariant. The key feature of a scale-free network is the long tail of the distribution. This means that there are nodes, albeit sparse, that have a large number of connections. In general, scale-invariance implies that there exists a fundamental relation that doesn't change when some independent variable is multiplied by constant. This is evident in equation \ref{PowerLaw}: if $k$ is scaled by an arbitrary constant, the only factor that changes is the normalization constant $C$. Scale invariance is a symmetry. Like any symmetry, it simplifies the attendant model of any physical systems it represents. 

Barab\'asi and Albert's {\em Science} article, with its emphasis on the scale-invariant symmetry of power law distributions, struck a chord with physicists primed by the simplicity of scale-invariant symmetries \cite{wilson1979problems}. They found that the scale-free property emerged from two biases. First, nodes are linked to existing nodes, consequently older nodes receive more connections. This first bias is a natural consequence of any evolving network where the growth consists of a steady addition of new nodes and their subsequent connections. Second, the new connections are attached to existing nodes with a probability proportional to the number of connections preexisting on a potential target. This latter bias they dubbed preferential-linking.  As preferential linking is probabilistically dependent on the degree of connections of the existing nodes, it implies that global knowledge is available and utilized in order to make new attachments. 

Barab\'asi and Albert were not the first to recognize the applicability of power law distributions to a wide assortment of natural phenomenon. In fact, Barab\'asi discusses the history of the mathematics and applications of power law distributions that preceded the 1999 paper in (\cite{Barabasi2016NetworkScience} pp. 116, 189). Particularly relevant are Herbert Simon's investigations of distributions, like power laws, that have long tails\cite{simon1955class}. Derek de Solla Price was perhaps the first to utilize preferential-linking in a  network model (\cite{price1965networks},\cite{price1976general}) of scientific citations, and he highlighted the power law behavior. Mark Newman presented a cogent and detailed re-examination of the Price model,  in  (\cite{mej2010networks} pp 487-513).

V\'{a}zquez \cite{PhysRevE.67.056104} and Nather, et al. \cite{nather2009hierarchy} have presented models of evolving networks with local rules. In \cite{nather2009hierarchy} Nather et al demonstrate a model that is scale-free and yields hierarchical clustering. Similarly, in the model by V\'{a}zquez, nodes are connected to previous nodes in the neighborhood of the new node. The networks generated are also scale-free and demonstrate a natural clustering hierarchy. In both of these models connections are made probabilistically to nearby nodes. Consequently they require knowledge of their neighborhood. In contrast, our intent is to eliminate $any$ requirement for knowledge of external topology. The only local connections in our network are the connection from the parent to a child and vice versa. 

\begin{figure}
\centering
\includegraphics[width=.5\textwidth]{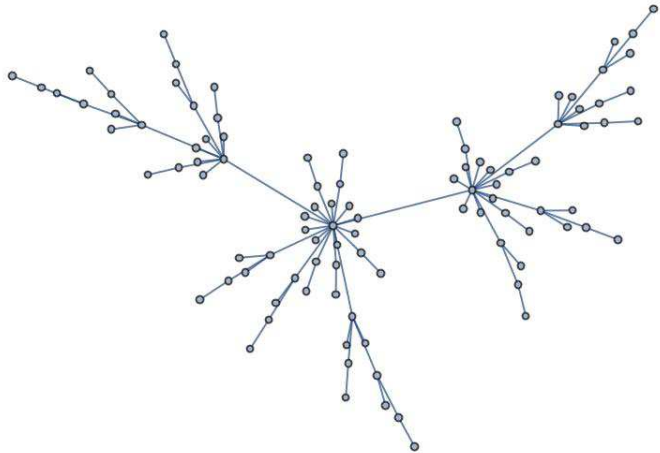}
\caption{A spawned network of one hundred nodes. The node with the most connections, roughly centered in the figure, is node 1, the seed node. The node just to its right is node 2.
\label{graphimage}}
\end{figure}

Here we present a local, deterministic, scale-invariant growth model that is neither probabilistic in how connections are made nor one that require global knowledge of the connection degree of any other nodes. This new model satisfies a different sort of preferential attachment. Nodes are spawned by existing nodes at a rate approximately proportional to the number of connections possessed by the parent and inversely proportional to time elapsed since the initial birth of the first node. Nodes with more connections spawn faster.  Our intent with this scheme was to eliminate any requirement for knowledge of external topology and still obtain scale-invariant symmetry. In particular, we were drawn to the idea that offspring could offer positive reproductive feedback. The only local connections in our network are the parent to child and vice versa.  Consequently, if the connection to an offspring provided additional reproductive energy, a parent's fecundity increases. Figure \ref{graphimage} shows such a constructed network of one hundred nodes. 

This idea of rapid unlimited nodal growth by nodes giving birth to new nodes was inspired by inflation models of the early universe. From an energy perspective, the $fuel$ for the enhanced growth in our model has to come from some infinitely large $nutrient$ rich environment. Similarly, In inflationary models, the energy density is fixed by a scalar field which provides the fuel for the expansion. This runs contrary to the usual conception that density is inversely proportional to volume. For a pedestrian view see \cite[p.~170]{guth1998inflationary})

Ours is not the only scale-free deterministic model. In \cite{barabasi2001deterministic}, Barab{\'a}si et al. demonstrate a clever copying scheme. As the network grows, patches of the network are copied and attached to the existing network.The resulting network has a degree exponent of $\alpha = \frac{\log(3)}{\log(2)}$. Obviously, this scheme requires knowledge of previous parts of the current network.

In the next section we present the model details and simulations. In section \uppercase\expandafter{\romannumeral3 \relax}, we discuss the theoretical underpinnings, in \uppercase\expandafter{\romannumeral 4\relax} we show the results of the simulations and compare them with theoretical predictions of the degree distributions, and in V we offer conclusions and a review. 

\section{\label{Simulation} model simulation}
We simulated the model using the methods outlined by Newman in \cite{mej2010networks} (pp. 282-290). The simulation software maintains the network by labeling every new node with an integer and notes whenever that node spawns. At time $t=1$ we assume that node 1 simply exists and is a starting seed for the network.  Further, we assume that node 1 birthed node 2 at $t=2$, and they are connected with an undirected link. Thus, the simulation begins at $t=3$ with the two connected nodes. Node 1 is initialized to spawn at $t=3$; node 2 at $t=4$. At each subsequent time step all existing nodes are examined to see if they are ready to spawn. Every node has an internal countdown timer. When a node is examined, and its timer is zero, the following occurs: 1) a node spawns a new node, 2) the parent's timer is reset to be current time divided by the number of connections of the parent minus one. 3) As it has no previous connections, the child's count-down timer is set to be proportional to the current time minus one. 4) an undirected edge is established between the parent node and offspring. 5) During this examination of all nodes, nodes that don't spawn have their timers decremented by one. After all the nodes are examined for the possibility of spawning, external time is incremented by one.

For example, consider the timeline of node 2 just after it spawns at time $t=4$. It has one prior connection---the connection from node 1. Therefore node 2's timer is set to $4/1 - 1=3$. It will spawn again when its timer decrements to zero.  Misleadingly, because the simulation is serialized, it requires four time steps until it spawns again. On the third subsequent time increment, at $t=7$, the timer steps down to zero. However, it is not discovered until the clock ticks one more time, and all nodes are examined to see if their timer is zero. So, node 2 will spawn again when $t=8$. 

\begin{figure}
\centering
\includegraphics[width=.6\textwidth]{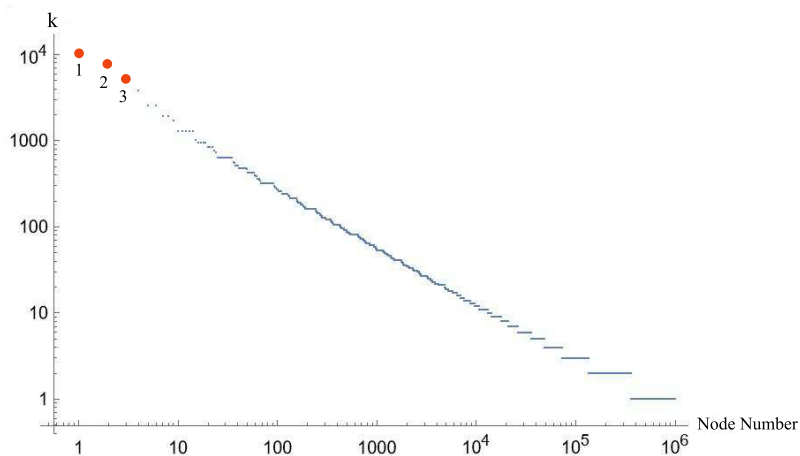}
\caption{Log-Log Plot of Number of Connections vs. Node Label. The 1st, 2nd, and 3rd nodes are highlighted in red.\label{networkgraphdisplay} }
\end{figure}

Figure \ref{networkgraphdisplay} shows the raw result of a simulation in the form of a log-log plot of the number of connections verses the node label. The simulation generated a total 1 million nodes. Since many nodes can be produced in a single external time step, only about 30,000 such steps are required to produce a network of this size.  

The deterministic nature of the model is evident by examining the spawning times of individual nodes. Let $T(n)$ be the time of a node spawning for the $n_{th}$ time. Since a node acquires another connection when it spawns, $n$ is also equal to the number of connections. Following the description of model, the relation between the time of the $n_{th}$ spawning and the previous spawning time is given by the recursion relation

\begin{equation}
     T(n)=T(n-1)+\frac{T(n-1)}{n-1}.
     \label{recursionrelation}
\end{equation}

 With the specification of the first spawning time, we can use Eq.(\ref{recursionrelation}) to find the time of spawning of any offspring of any node. As an example we can compute the $4_{th}$ spawning time of the second node. The initial spawning time of node 2 is 4, so $T(1)=4$. Applying Eq(\ref{recursionrelation}) once gives the second spawning time 8, twice gives 12, and three times 16, thus $T(4)=16$. In fact, if $T(1)=T_0$ is the first spawning time of a node, then it is readily seen that $T(n)=nT_0$ is the $n_{th}$ spawning time.

\section{\label{Theory}Theory}

In this section, we leverage the theory behind the model developed by Derek de Solla Price in the 1970s \cite{price1976general} and in Newman's master equation alternative analysis to the Price model in $Networks$  (\cite{mej2010networks} pp. 487-495). Price's picturesque analogy of negatively decaying atoms mirrors our spawning nodes. We also leaned on Newman's analysis of the Yule process in \cite{Newman2006Power-Laws}. 
 
Let $n$ be the total number of nodes in the network at the current time. In the usual way we separate the overall collection of nodes into non-overlapping sets.  Every node in a set has the same number of connections. Different sets have nodes with a different number of connections, $q$. By the nodes spawning, the set  with $q$ connections feeds the set of nodes with $q+1$ connections, while gaining nodes from the set with $q-1$. Whenever any node spawns in any of the sets, the child enters the set of $q=1$. This $q=1$ set loses members to the $q=2$ set whenever any of its members spawn. Because it contains all the most recent offspring, this set will always have the most members. Subsequently, we will develop an expression of the $degree$ $distribution$: the fraction of all nodes in each set. Let $p_{q}(n)$ be that distribution. 
 
Consider the case when a single node in any set spawns. First note that the population of the entire network increases by one. Second, a spawning parent automatically receives a connection from its offspring. Consequently, a parent node will move from a set of nodes with $q$ connections to the set of nodes with $q+1$. As we discussed in the previous section, the time interval before the next spawning is approximately proportional to the current time divided by the number of connections. Therefore, the $spawning$ $rate$, the inverse of that time interval, will be approximately proportional to the number of connections divided by the current time.  Any node with the same number of connections is equally likely to spawn in some short time interval.  The more connections a node has, the higher the likelihood of it spawning. However, the probability of a node spawning cannot be strictly proportional to the number of connections. (This issue is discussed at length in \cite{mej2010networks} and in  \cite{price1976general}.) The probability of a spawning must be proportional to the number of connections plus some additive constant: $q+a$. In our case $a=1$.) To normalize the probability of a node spawning, we divide by the sum over all the nodes in the network plus one.  Thus, the probability of a node spawning is 
\begin{equation}
     \frac{q+1}{\sum_{i}^n (q_i+1)} = \frac{q+1}{\sum_{i}^n q_i + \sum_{i}^n 1}.
     \label{SpawningProbability}
\end{equation}
We can simplify this result by utilizing the average number of connections per node. Let $Q$ be the total number of connections for all nodes. From our simulations, with $N=1,000,000$, we have $Q=\sum_{i=1}^{N}q_i=1,999,998$. So, $<q>=2$. Eq.(\ref{SpawningProbability}) then becomes
\begin{equation}
    \frac{q+1}{2n + n}.
    \label{SimplifiedSpawingProbability}
\end{equation}
 \noindent The denominator of Eq.(\ref{SimplifiedSpawingProbability}) then becomes $3n$ and the probability of a node spawning is 
\begin{equation}
     \frac{q+1}{3n}.
     \label{SpawningProbability2}
\end{equation}

If the average number of connections per node is two, then the anticipated number of new connections due to nodes spawning is

\begin{equation}
     2\frac{q+1}{3n}.
     \label{AnticipatedNumber}
\end{equation}

From the definition of $p_{q}(n)$ as the fraction of nodes that have q connections, we see that there are $np_q(n)$ nodes within a set of nodes with q connections.  Consequently, the assumed number of spawnings within some small time interval for nodes with $q$ connections is 

\begin{equation}
     np_{q}(n)2\frac{(q+1)}{3n}=\frac{2(q+1)}{3}p_{q}(n).
     \label{NumberOfSpawns}
\end{equation}

Reiterating: when parent nodes with $q-1$ connections spawn they move from the set of nodes with $q-1$ connections to the set of nodes with $q$ connections. Therefore, the expected number of nodes that make this jump and add to the set of q connections is  

\begin{equation}
    \frac{2}{3}qp_{q-1}(n).
     \label{JumpsInto}
\end{equation}

\noindent Similarly, the expected number of nodes in the set with $q$ connections that will be lost will be  

\begin{equation}
    \frac{2(q+1)}{3}p_{q}(n).
     \label{JumpsOut}
\end{equation}

The change in the number of nodes for the set with nodes of $q$ connection is then the number of nodes that jump into the set minus the number of nodes that jump out of the set: 

\begin{equation}
    (n+1)p_{q}(n+1)-np_{q}= \frac{2}{3}qp_{q-1}(n)-\frac{2(q+1)}{3}p_{q}(n).
     \label{MasterEquation}
\end{equation}

Eq.(\ref{MasterEquation}) is the master equation for our spawning process. It is valid for all values of $q$ greater than 1. By first letting $n\rightarrow\nolinebreak\infty$, we iterate Eq.(\ref{MasterEquation}) with an appropriate initial value for $p_1$ to generate an asymptotic form of $p_q$. (Notionally we will indicate this transition to large $n$ by simply writing $p_q$ instead of $p_q(n)$.) 

For the initial case of $q=1$, we gain a child node when any node in the network spawns, thus the first term on the right in Eq.(\ref{MasterEquation}) is simply one. The initial condition is then 

\begin{equation}
    p_1=1-\frac{4}{3}p_1,
     \label{InitialCondition1}
\end{equation}
or
\begin{equation}
    p_1=\frac{3}{7}.
     \label{NewInitialCondition}
\end{equation}

\noindent We can also rearrange and simplify Eq.(\ref{MasterEquation}) to get

\begin{equation}
    p_{q}=\frac{q}{\frac{5}{2}+q}p_{q-1}
     \label{MasterEquation2}
\end{equation}

We iterate Eq(\ref{MasterEquation2}) starting with $p_1=\frac{3}{7}$. Two constants, $\mu = \frac{7}{2}$ and $\delta = \frac{3}{7}$, facilitates recognizing and simplifying patterns. The resulting expression after iterating is:

\begin{equation}
    p_q=\frac{\delta(1+1)(1+2)(1+3)\cdots(1+(q-1))}{\frac{1}{\mu}(\mu+1)(\mu+2)(\mu+3)\cdots(\mu+(q-1)}.
     \label{Iteration1}
\end{equation}

\noindent Gamma functions with integer arguments are often associated with expression like that in  Eq.(\ref{Iteration1}). For example, we can use the identity below, Eq.(\ref{GammaIndent}) (see \cite{abramowitz1965handbook} pg. 256),
\begin{equation}
x(x+1)(x+2)\cdots(x+n-1)=\frac{\Gamma(x+n)}{\Gamma(x)},
\label{GammaIndent}
\end{equation}
to transform and simplify Eq.(\ref{Iteration1})  into expressions involving gamma functions. With this identity equation \ref{Iteration1} becomes 

\begin{equation}
\mu\:\delta\:\frac{\Gamma(1+q)\Gamma(\mu)}{\Gamma(\mu+q)},
\label{IterationGamma}
\end{equation}

\noindent where we have used $\Gamma(1)=1.$

The asymptotic form we seek is easier to find if we convert some of the gamma functions to beta functions using the identity,
\begin{equation}
\Gamma(x+y)=\frac{\Gamma(x)\Gamma(y)}{\mathrm{B}(x,y)}.
\label{GammasAndBetas}
\end{equation}
Applying Eq.(\ref{GammasAndBetas}) to Eq.(\ref{IterationGamma}) yields,

\begin{equation}
\mu\:\delta\:\frac{\mathrm{B}(\mu,q)}{\mathrm{B}(1,q)}.
\label{AlphasIntoBetas}
\end{equation}

When $x$ is large and $y$ is fixed, the asymptotic form of the beta function is $\mathrm{B}(x,y)\sim\Gamma(y)x^{-y}$ . Using this result in Eq.(\ref{AlphasIntoBetas}) with $q$ large, we get 

\begin{equation}
\mu\:\delta\:\Gamma(\mu)\:q^{1-\mu}.
\label{AlmostThere}
\end{equation}

\noindent Substituting the values of the constants, and $\Gamma(\frac{7}{2})=\frac{15\sqrt{\pi}}{8}$ into Eq.(\ref{AlmostThere}), we obtain the asymptotic form for $p_q$,

\begin{equation}
p_q\sim\frac{45\sqrt{\pi}}{16}\:q^{-\frac{5}{2}}.
\label{There}
\end{equation}

\noindent For large values of $q$ Eq.(\ref{There}) shows the classic power law behavior with $\alpha=2.5$.  

\section{\label{Results} Results}
The simulation aggregates nodes with the same number of connections into sets. Consequently, we can explicitly examine the growth of these sets. In particular, we study their overall growth rate and how set membership changes per unit time.The simulation shows that the number of nodes from all sets increases slowly initially then accelerates obeying the power law,
  
\begin{equation}
    y=0.27t^{1.46}.
\label{PowerLawGrowth}
\end{equation}

With successively decreasing importance, nodes with the smallest number of connections dominate overall network growth. For instance, nodes with just one connection provide about 60\% of the nodes out of a total of $1,000,000$. Nodes with two connections provide 7\% of the total. Since the overall network growth obeys a power law, it is not surprising that the nodal growth of individual sets of nodes with the same number of connections show similar behavior. For instance, growth for nodes in the set with two connections obey the power law 

\begin{equation}
y=0.018t^{1.47}.
\label{PowerLawGrowthForSet2}
\end{equation}

\begin{figure}
\centering
\includegraphics[width=.6\textwidth]{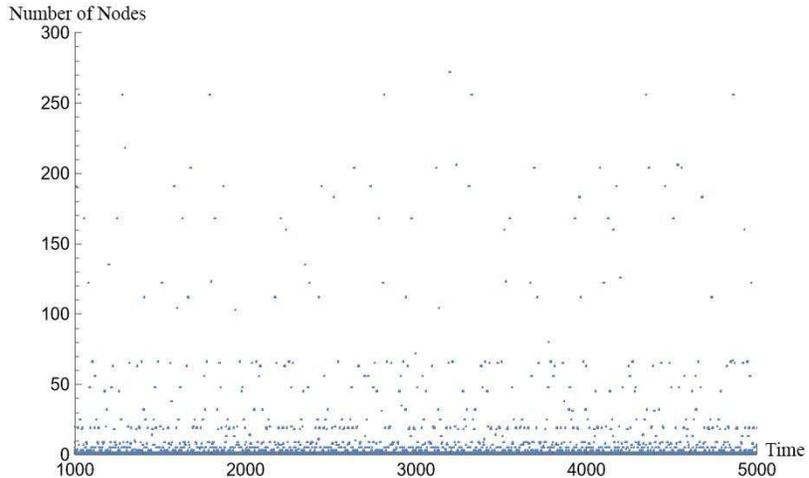}
\caption{Number of nodes spawned per unit time from $t=1000$ to $t=2000$.
\label{randomness} }
\end{figure}

\begin{figure}
\centering
\includegraphics[width=.6\textwidth]{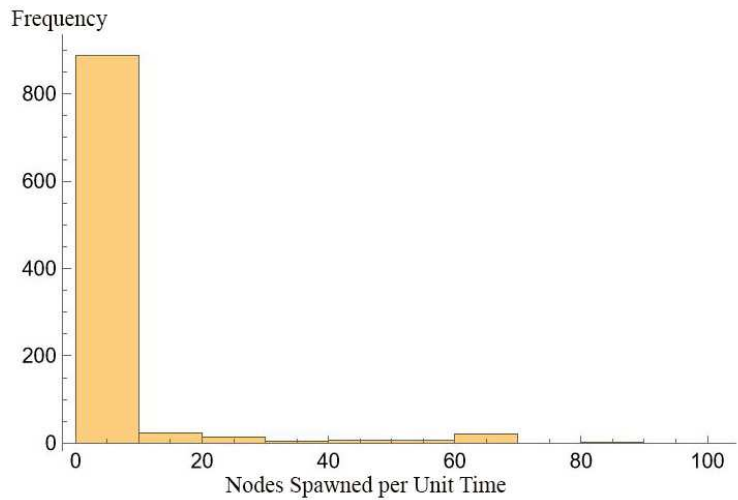}
\caption{Histogram of the frequency of births of nodes per unit time.}
\label{histogram} 
\end{figure}

Even though our model is deterministic, the results can be quasi-random. Unlike the behavior of some non-linear dynamical systems, where chaotic behavior is caused by sensitivity to initial conditions, the randomness in our model is a consequence of the inter-generational mixing of birthing times. Because nodes are likely to have a different number of connections, the timing of new births intermingle. For example, four new nodes are born at $t=54$.  Node 1 gives birth to its $18$th offspring, node 94; node 3 gives birth to its $9$th offspring, node 95; node 9 gives birth to its 4th offspring, node 96; and node 37 gives birth to its second offspring, node 97. However, at $t=53$ their are no births. Consequently, without the maintenance of a detailed log of when every node is born and its spawning times, as the computer code provides, it is difficult to predict when a particular time step has no births, or 260 births. Figure \ref{randomness} offers a glimpse of this noise like behavior. It shows the number of nodes produced during each unit time step over the span of $t=1000$ to $t=5000$. In this time range the number of nodes produced in a unit time step vary from 0 to 260. We also generated an empirical distribution. (An empirical distribution is a non-parametric way of establishing a distribution-like object based on empirical data.) of the number of nodes produced in 30,000 time steps\textbf{\textemdash}the approximate number of time steps required to produce $1,000,000$ nodes. The distribution has a mean of 32 and a standard deviation of 440 nodes spawned. Figure \ref{histogram} shows a histogram of 1000 random deviates taken from the this distribution. Note that histogram shows a long tail. Unfortunately the distribution is not characterized simply by known distributions.

\begin{figure}
\centering
\includegraphics[width=.6\textwidth]{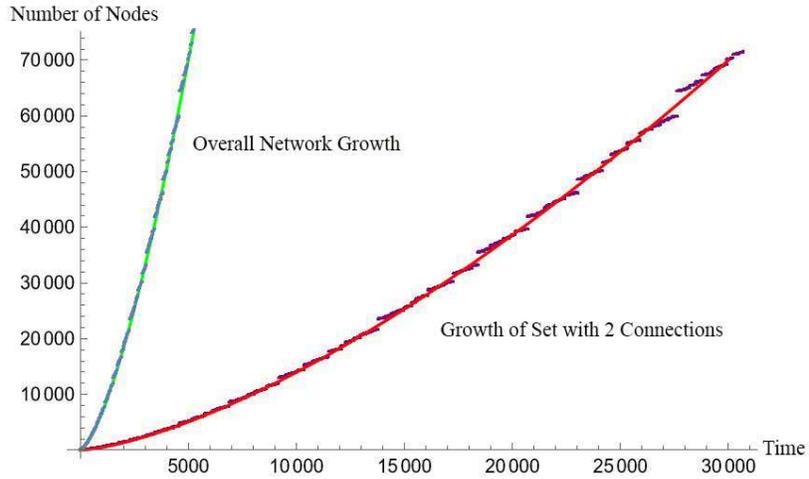}
\caption{Network Growth. Overall network growth shown in dotted blue with power law curve fit in solid green. Growth of nodes in the set with two connections is shown in dotted purple with power law curve fit in solid red line. (The growth of the total number of nodes looks as though it would exceed 1,000,000 nodes in much less time than is actually the case. This is misleading. For $y=1,000,000$ and solving for time in Eq.(\ref{PowerLawGrowth}) yields $x=31,554$ which is consistent with our simulations.)}
\label{combinedgrowth} 
\end{figure}

The simulation provides a list of the number of connections for each node. Recall that each node is labeled in the order it was produced. The older nodes always have more connections. Therefore the list is automatically sorted by the decreasing number of connections. This was shown in the log-log graph in figure \ref{networkgraphdisplay}. The negative sloping straight line of this graph strongly hints that the network's degree distribution obeys a power law. We generated another empirical distribution from this list and took random variates to examine the resulting degree distribution.    

Given this automatic sorting of the most connected nodes to the least connected, it would seem that the Zipf distribution is a natural choice for the resulting distribution. The inception of Zipf's law comes from the observation by George Zipf that if words in any written language are sorted by the most frequently used to the least used their ranking follows a power law. (\cite{schroeder2009fractals} pp. 35-37) More formally, the Zipf distribution is a discrete distribution with a long tail. Equation \ref{Zipf} shows its probability mass function. 

\begin{equation}
f(x)=\frac{x^{-(\rho + 1})}{\sum_{i=1}^{n}x_{i}^{(\rho+1)}},
\label{Zipf}
\end{equation}

\noindent where $\rho$ is a parameter greater or equal to zero and n is a natural number. 

\begin{figure}
\centering
\includegraphics[width=.6\textwidth]{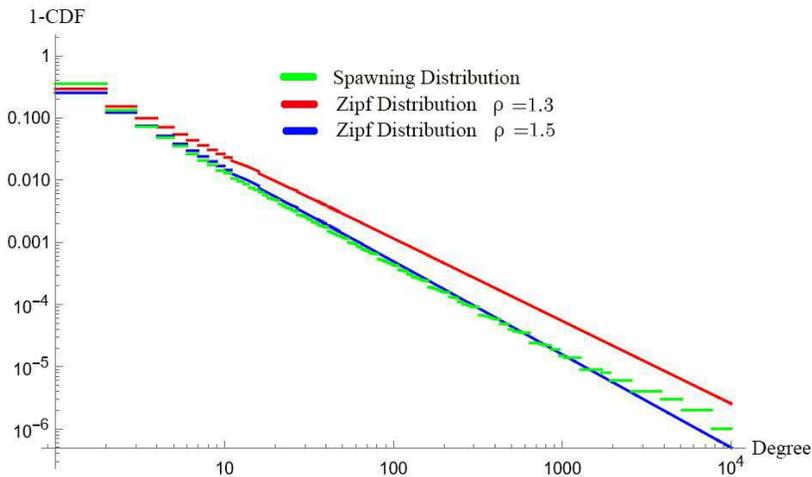}
\caption{Cumulative Distributions of the Spawning Model and the Zipf Distribution with parameter 1.5.
\label{networkdistribution} }
\end{figure}

Assuming the results indicated we had degree distribution with a power law of the form Eq.(\ref{PowerLaw}), we attempted the methods discussed in \cite{Clauset2009Power-Law} to identify the value of $\alpha$. These methods require that we omit some of initial values of the data with a parameter $x_{min}$. However, we found the methods unstable with respect to any value of this parameter. We also attempted optimizing the fit of the Zipf distribution parameter to the data using different statistical methods and even different discrete distributions with long tails. These methods consistently pointed to a Zipf distribution with a value for $\rho=1.32$. However, with this value of $\rho$ there was a clear difference in a simple visual inspection on a log-log plot of one minus the cumulative probability distribution (CDF) of samples of the Zipf distribution and $1-CDF$ of samples of the empirical distribution of our results. We are aware of the problems of relying on visual fits, nevertheless we believe that the best value of $\rho$ is 1.5. From the probability mass function Eq.(\ref{Zipf}) of the Zipf distribution, we see that a value of $\rho=1.5$ implies a power law distribution with an exponent of $\alpha=2.5$. This is consistent with our results in \uppercase\expandafter{\romannumeral 3\relax} where we predicted a power law with $\alpha=2.5$. The fit it offers we show in Eq.(\ref{networkdistribution}). In that figure we also show the Zipf distribution with a parameter $\rho=1.3$.

\section{\label{conclusion}Conclusion}
We have offered a new deterministic scale-invariant description of a unique birthing process and the resulting network. This is a network formed solely by undirected connections of parents to their children. It emerges from the development of long chains of generations of offspring and the children of those offspring. The older the nodes, the longer the chains, and the larger the sub-network of offspring. The network develops without information being shared between nodes. The only requirement is that parents modulate their fecundity by making the time between births be approximately proportional to the current time divided by the number of previous births. The power law aspect of our model results simply from the requirement that nodes with more previous births spawn more often. 

Nodes must be able to maintain an internal clock in order to set their spawning time. It is likely that the internal timers will need to be synchronized. However, it is not clear to what extent the various features of the model breakdown when the internal clocks between nodes deviate significantly. It is possible that we have introduced unanticipated artifacts because of the serial nature of the simulation. Conceivably, this could be addressed by randomizing the inspection of node timers for each time step. These explorations we save for future work. Also, we attribute the issues we had with standard statistical methods to estimate the power law exponent to be due to the unique features of our model. Our hypothesis is twofold. First, we have a very small number of nodes with many connections. These make the statistics unreliable for this part of the data. Second, whenever a simulation is halted, be it for ten thousand nodes or one million nodes it is a hard stop. There is always a large bolus of nodes on the verge of birthing in the next time increment. We think this biases the statistics for nodes with few connections. This too needs to be investigated further.  

The authors acknowledge the help of John Hurdle MD,PhD in their preparation of this document.

\newpage
\bibliographystyle{plain}
\bibliography{graph.bib}
\end{document}